\documentclass[]{spie}  

\usepackage{amsmath,amsfonts,amssymb}
\usepackage{graphicx}
\usepackage[colorlinks=true, allcolors=blue]{hyperref}
\usepackage{float}
\usepackage{wrapfig}

\usepackage{listings}
\usepackage{xcolor}

\definecolor{codegreen}{rgb}{0,0.6,0}
\definecolor{codegray}{rgb}{0.5,0.5,0.5}
\definecolor{codepurple}{rgb}{0.58,0,0.82}
\definecolor{backcolour}{rgb}{0.95,0.95,0.95}

\lstdefinestyle{mystyle}{
	backgroundcolor=\color{backcolour},   
	commentstyle=\color{codegreen},
	keywordstyle=\color{magenta},
	numberstyle=\tiny\color{codegray},
	stringstyle=\color{codepurple},
	basicstyle=\ttfamily\footnotesize,
	breakatwhitespace=false,         
	breaklines=true,                 
	captionpos=b,                    
	keepspaces=true,                 
	numbers=left,                    
	numbersep=5pt,                  
	showspaces=false,                
	showstringspaces=false,
	showtabs=false,                  
	tabsize=2
}

\title{Poke: an open-source ray-based physical optics platform}

\author[a]{Jaren N. Ashcraft}
\author[b]{Ewan S. Douglas}
\author[a,b,c]{Daewook Kim}
\author[d]{A.J. E. Riggs}
\author[b]{Ramya Anche}
\author[a]{Trent Brendel}
\author[a]{Kevin Derby}
\author[d]{Brandon D. Dube}
\author[a]{Quinn Jarecki}
\author[a]{Emory Jenkins}
\author[a]{Kian S. Milani}
\affil[a]{James C. Wyant College of Optical Sciences, University of Arizona, Meinel Building 1630 E. University Blvd., Tucson, AZ. 85721, USA}
\affil[b]{Department of Astronomy and Steward Observatory, University of Arizona, 933 N. Cherry Ave., Tucson, AZ 85721, USA}
\affil[c]{Large Binocular Telescope Observatory, University Of Arizona, 933 N. Cherry Ave. Tucson, AZ 85721, USA} 
\affil[d]{Jet Propulsion Laboratory, California Institute of Technology, 4800 Oak Grove Drive, Pasadena, CA 91109, USA}

\authorinfo{Further author information: (Send correspondence to J.N.A)\\J.N.A.: E-mail: jashcraft@arizona.edu}

\begin{document} 
\maketitle

\begin{abstract}
Integrated optical models allow for accurate prediction of the as-built performance of an optical instrument. Optical models are typically composed of a separate ray trace and diffraction model to capture both the geometrical and physical regimes of light. These models are typically separated across both open-source and commercial software that don’t interface with each other directly. To bridge the gap between ray trace models and diffraction models, we have built an open-source optical analysis platform in Python called Poke that uses commercial ray tracing APIs and open-source physical optics engines to simultaneously model scalar wavefront error, diffraction, and polarization. Poke operates by storing ray data from a commercial ray tracing engine into a Python object, from which physical optics calculations can be made. We present an introduction to using Poke, and highlight the capabilities of two new propagation modules that add to the utility of existing scalar diffraction models. Gaussian Beamlet Decomposition is a ray-based approach to diffraction modeling that allows us to integrate physical optics models with ray trace models to directly capture the influence of ray aberrations in diffraction simulations. Polarization Ray Tracing is a ray-based method of vector field propagation that can diagnose the polarization aberrations in optical systems. Poke has been recently used to study the next generation of astronomical observatories, including the ground-based Extremely Large Telescopes (TMT, GMT, ELT) and a 6 meter space telescope (6MST) early concept for NASA’s Habitable Worlds Observatory.
 
\end{abstract}


\section{INTRODUCTION}
\label{sec:intro}  

Optical modeling is an integral engineering tool in instrument development. In imaging applications, the variety of optical modeling tools used by the industry typically fall into two categories:
\begin{itemize}
    \item \emph{Ray Tracers} (e.g. CODE V\footnote{\url{https://www.synopsys.com/optical-solutions/codev.html}}, OpticStudio\footnote{\url{https://www.ansys.com/products/optics-vr/ansys-zemax-opticstudio}}) are used to optimize and tolerance the design of optical instruments. We use ray tracers to optimize the shapes of lens or mirror surfaces to minimize the scalar wavefront aberrations that limit image quality.
    \item \emph{Physical Optics Propagation Packages} (e.g. POPPY\cite{2016ascl.soft02018P}, HCIPY\cite{HCIPYdocs}) are used in the design and modeling of systems that contain elements that can't be represented by a ray trace, such as diffractive masks or surface aberrations from polishing errors. 
\end{itemize}

\emph{Ray Tracers} are one of the cornerstones of the optical engineering industry. Their ability to simulate and design real optical systems for accurate performance predictions makes them a necessity in any optical modeling pipeline. However, ray tracers have a limited capacity to perform diffraction simulation and the proprietary ray tracers (e.g. CODE V, Zemax) are more commonly used than some of the excellent packages in the open-source\footnote{ray-optics \url{https://github.com/mjhoptics/ray-optics}} \footnote{raypier \url{https://github.com/bryancole/raypier_optics}}. This limits the use of industry-standard ray tracers to those who are able to afford a license. 

Open-source \emph{Physical Optics Propagation Packages} on the other hand are not as restricted. They are common in fields that require modeling of an optical system in the diffraction limit. Many were originally developed to support astronomical high-contrast imaging (e.g. POPPY, HCIPy, prysm\cite{Dube2019}), but have been used in other fields since their inception. Their free and open source nature means that they are accessible to everyone and the algorithms used are transparent.

There is presently a disparity in the capabilities of open-source modeling packages and commercial ray tracers (see Table \ref{tab:capabilities}). A unified optical modeling pipeline would have the capability to perform ray traces, plane-to-plane diffraction simulation, and ideally be free and open source to ensure accessibility. However, the ubiquity of commercial ray tracers mean that models must be able to extract data from them. While the utility of commercial ray tracers is undeniable, it results in a permanent disconnect between ray tracers and open-source diffraction modeling packages.

\begin{table}[H]
    \centering
    \begin{tabular}{c c c c c c}
        \hline
        Code & Free and Open-Source & Ray tracing & PSF simulation & HCI toolkit & Polarization  \\
        \hline
        CODE V & & \checkmark & \checkmark & & \checkmark  \\
        ZEMAX & & \checkmark & \checkmark & & \checkmark   \\
        FRED & & \checkmark & \checkmark & & \checkmark    \\
        POPPY & \checkmark & & \checkmark & \checkmark &  \\
        HCIPy & \checkmark & & \checkmark & \checkmark & \checkmark \\
        prysm & \checkmark & experimental & \checkmark & \checkmark & experimental \\
        \hline
        \hline
        
    \end{tabular}
    \caption{Sample optical modeling codes and their capabilities. HCI toolkit means a framework for modeling the diffraction from occulters and surface errors with high accuracy. Note that the prysm optical propagation package has some limited ray tracing and polarization utility in development.}
    \label{tab:capabilities}
\end{table}

Fortunately, the most popular commercial ray tracers have APIs that allow the user to interface with the raytracer using a programming language. Common among these is a Python API. Python is one of the most popular programming languages, particularly among scientists. Given the industry's general fluency in the language, a software platform built in Python that interfaces with commercial ray tracing APIs \emph{and} physical optics propagators has the potential to unite the disparate modeling regimes.

\begin{wrapfigure}{R}{0.22\textwidth}
\centering
\includegraphics[width=0.2\textwidth]{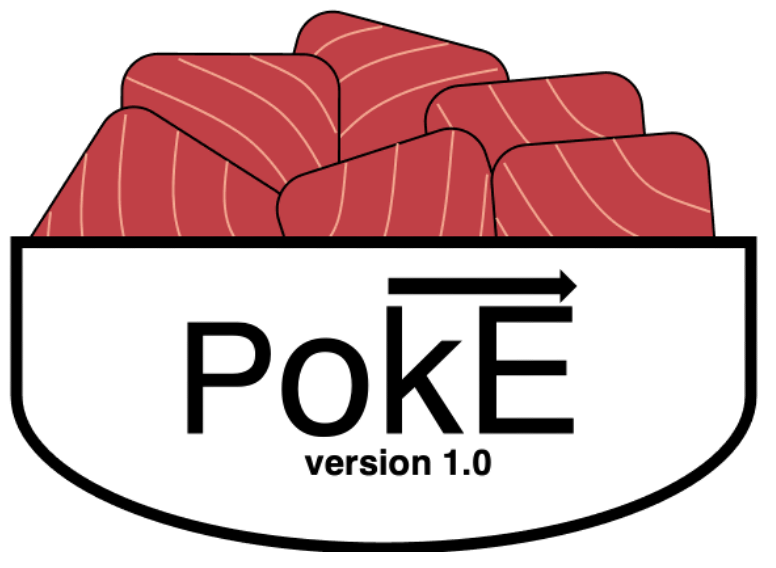}
\label{fig:pokelogo}
\caption{Poke Logo}
\end{wrapfigure}
Poke (logo in Figure \ref{fig:pokelogo}) is a Python 3.8+ package currently on Github\cite{jaren_ashcraft_2023_7951685} that aims to accomplish this goal. Poke was originally built as a polarization ray tracing (PRT) engine to simulate the influence of polariazation aberrations in astronomical telescopes \cite{Anche2023}. We've recently generalized Poke to be compatible with the Zemax and CODE V ray tracers, and several physical optics propagation packages. By doing so, we anticipate that Poke can serve as a Pythonic bridge from ray tracers to diffraction modeling packages, alleviating the need for the user to learn commercial ray tracing Python API's. In this manuscript we detail the basics of using Poke, and explore some of its current features.

In section \ref{sec:poke_core} we discuss the core functionality of Poke. In section \ref{sec:poke_physics} we discuss the two ray-based physical optics modules developed in Poke to enhance integrated modeling of diffraction and polarization. In section \ref{sec:poke_integrated} we showcase how Poke can be used to create integrated ray and wave based optical models. In section \ref{sec:discussion} we outline a path for Poke's continued development.

\newpage
\section{The Rayfront Interface}
\label{sec:poke_core}

Poke's sole interface utilizes the \verb|Rayfront| object, which contains the totality of the ray information needed for physical optics calculations. Rayfront is a portmanteau of ``Ray" and ``Wavefront", encapsulating Poke's mission to link ray tracing with wave propagation. The \verb|Rayfront| is first established by initializing it with parameters of the optical system, such as number of rays, wavelength, aperture, and field of view (shown below). 

\begin{lstlisting}[language=Python, caption=Generation of a Rayfront object. This is the basic interface required to use Poke.]
from poke.poke_core import Rayfront

# parameters needed to setup Rayfront
pth_to_lens = 'hubble_test.len'
number_of_rays = 20 # across the entrance pupil
wavelength = 1e-6  # meters
pupil_radius = 1.2 # semi-aperture of Hubble
max_field_of_view = 0.08 # degrees

rf = Rayfront(number_of_rays,wavelength,pupil_radius,max_field_of_view,circle=True)
\end{lstlisting}

We then define a list of surfaces that we want our \verb|Rayfront| to save data for. In the example above, we care about the primary and secondary mirror. The surface data required of the user is specified using Python dictionaries, which acts like a low-level ``user interface" to set up optical system data per-surface.
The \verb|trace_raysets| method is called when we are ready to trace rays, and it is currently compatible with both CODE V and Zemax ray tracing packages. Running this method will run a ray trace dependent on the file extension of the lens path specified, and save the important ray data to the \verb|Rayfront|.
\begin{lstlisting}[language=Python, caption=Running a raytrace through the CODE V lens file  by calling the trace raysets method. This method works works for both CODE V and Zemax files]
s1 = {'surf': 1, 'coating': 1 + 1j*7, 'mode': 'reflect'}
s2 = {'surf': 2, 'coating': 1 + 1j*7, 'mode': 'reflect'}

rf.surfaces =[s1,s2]
rf.trace_raysets(pth_to_lens)
\end{lstlisting}

\verb|Rayfront|s can be written to and read from a binary file using the \verb|msgpack| Python package which provides efficient data storage and fast serialization. Simply call the appropriate function from the writing module.
\begin{lstlisting}[language=Python, caption=A Rayfront can be written to a file using the poke.writing module. It can then be distributed and read back into Poke effectively open-sourcing the ray data.]
from poke.writing import read_rayfront_from_serial,write_rayfront_to_serial
write_rayfront_to_serial(rf,'sample_rayfront.msgpack')
rf = read_rayfront_from_serial('sample_rayfront.msgpack')
\end{lstlisting}
The read/write capability of the \verb|Rayfront| provides a compact and useful way of distributing ray data that is strictly agnostic of the ray tracing engine used to generate it. This feature is core to the identity of Poke, because it effectively open-sources ray traces of optical systems. As long as one person in a given workforce can generate a \verb|Rayfront|, it can be distributed to any interested investigator. Permanently saving the ray data also allows the user to save a ``state" of the optical system, alleviating the need to re-run ray traces to perform analyses. The \verb|Rayfront| can also hold multiple sets of rays, meaning that a raytrace can be performed for multiple field points, wavelengths, or misalignments.

\subsection{Rayfront Attributes}
From the \verb|Rayfront| we have immediate access to data that describes the constituent rays. These attributes are shown in Table \ref{tab:attributes}.

\begin{table}[H]
    \centering
    \begin{tabular}{c c}
    \hline
         Rayfront.attribute & Description  \\
    \hline
         xData & ray position coordinate in x dimension \\
         yData & ray position coordinate in y dimension\\
         zData & ray position coordinate in z dimension\\
         lData & ray direction cosine coordinate along x-axis\\
         mData & ray direction cosine coordinate along y-axis\\ 
         nData & ray direction cosine coordinate along z-axis\\ 
         l2Data & surface normal direction cosine coordinate along x-axis\\
         m2Data & surface normal direction cosine coordinate along y-axis\\ 
         n2Data & surface normal direction cosine coordinate along z-axis\\ 
         opd & Optical path length the ray experienced \\
         \hline
    \end{tabular}
    \caption{Table of basic attributes of the Rayfront object.}
    \label{tab:attributes}
\end{table}

\begin{figure}[H]
    \centering
    \includegraphics[width=0.7\textwidth]{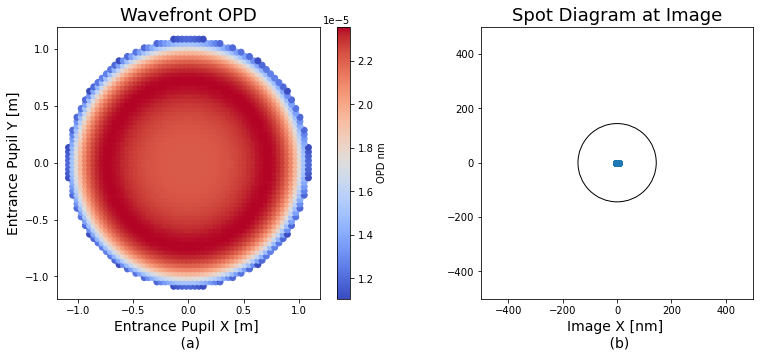}
    \caption{Illustration of the attributes accessible in Python via Poke's Rayfront. (a) A plot of the OPD map generated the previous code in Listing 4. (b) The spot diagram at the image plane, the diffraction-limited radius for this optical system is plotted as a black circle.}
    \label{fig:opd_spotsize}
\end{figure}

These data can be accessed directly from the \verb|Rayfront| object. They are \verb|numpy| arrays whose axes correspond to rayset, surface, and then ray (see the example in Listing 4). With the basic attributes, we can perform simple analyses such as spot diagrams or examining the OPD of the ray bundle at the exit pupil (shown in Figure \ref{fig:opd_spotsize}. 

\begin{lstlisting}[language=Python, caption= Visualizing an OPD map over the entrance pupil of the Hubble space telescope.]
rayset = 0
surf_xy = 0 # primary mirror
surf_opd = -1 # the image plane
x,y = rf.xData[rayset][surf_xy],rf.yData[rayset][surf_xy]
opd = rf.opd[rayset][surf_opd]
plt.scatter(x,y,c=opd,cmap='coolwarm')
\end{lstlisting}

We can access the coordinates at any surface in the optical system, which means that the totality of ray-based analyses can be built in Poke. However, Poke's primary utility is to serve as a platform for ray-based physical optics. In the next section we illustrate two key physics modules built into Poke that can simulate diffraction and polarization from purely ray data.

\section{Physical Optics Modules}
\label{sec:poke_physics}

Poke was developed with the intention of adding more open-source propagation physics to existing diffraction modeling packages. The two primary physics packages that use ray data are Gaussian Beamlet Decomposition and Polarization Ray Tracing.

\subsection{Gaussian Beamlet Decomposition}
Traditionally, physical optics propagators employ diffraction integrals derived from the Huygens-Fresnel principle. Plane-to-plane propagation is done with the Fresnel diffraction integral typically by using the angular spectrum method to model the transfer function of free space using a Fast Fourier Transformation. However, using these diffraction integrals imposes the scalar and paraxial assumption on the optical system of interest. This limits the scope of what the diffraction model can accurately determine.

\begin{figure}[H]
    \centering
    \includegraphics[width=\textwidth]{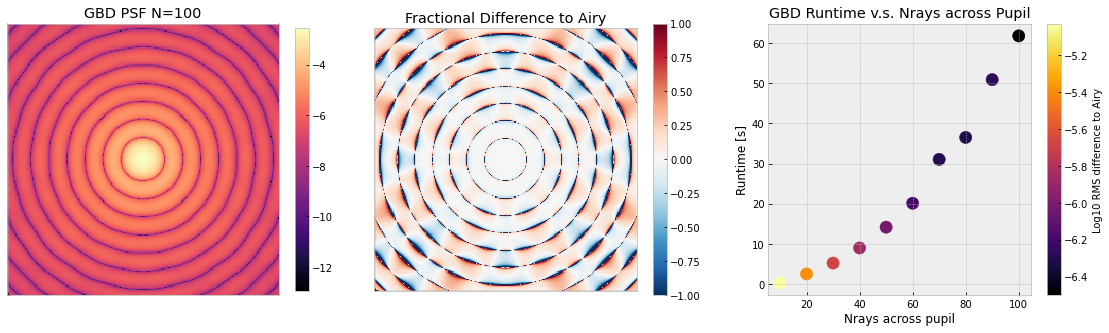}
    \caption{Comparison of a GBD PSF from generated by the Hubble Space Telescope with an unobscured aperture. (Left) is the PSF produced by GBD, (Middle) is the fractional difference to the analytic Airy function for the same aperture. (Right) shows how the computation time scales with the number of rays (or beamlets) used across the pupil and how accurate the results are. The computation complexity is roughly proportional to $N_{rays}^{2}$. The color of the spots show the Log10 RMS difference to the analytical Airy function, showing that more beamlets means a more accurate diffraction computation.}
    \label{fig:gbd_psf}
\end{figure}

Gaussian Beamlet Decomposition (GBD) is an alternative method of physical optics propagation that has seen substantive implementation in commercial software. The operating principle of GBD is to decompose an arbitrary field into a finite sum of coherent Gaussian beamlets. The analytical propagation laws for these beamlets are known via ray transfer matrices, so they can be propagated along geometric ray paths. GBD doesn’t have an apparent presence in the open source. This is likely because a clear algorithm for its implementation is not present in the literature. We derived a method for its implementation and built it into Poke to study its suitability for high-contrast imaging\cite{ashcraft_inreview}. The operating principle of GBD is exhaustively covered in the literature \cite{Harvey15,Worku19,Greynolds86}. Because we know the analytic propagation laws of Gaussian beams, we can decompose an arbitrary wavefront into a finite sum of Gaussian beams and propagate them anywhere in a 3D non-paraxial optical system. In Figure \ref{fig:gbd_psf} we show the ability of our \verb|poke.gbd| module to reconstruct the analytical airy function. GBD requires that each beamlet is propagated to each point that we wish to know the field at. If we desire to propagate a field decomposed into $N_{beamlets}$ to a plane that is $N_{pix} \times N_{pix}$ in size, then the total number of propagations is equal to $N_{beamlets} \times N_{pix} \times N_{pix}$. This is a daunting challenge computationally, so careful consideration was built into the structure of our algorithm. All computation is vectorized, with the option to break it the simulation up into loops over the beamlet index should the simulation require more memory than a given computer has. We have also implemented accelerated computing options for computation on GPUs.

Optimizing the GBD technique is an active area of research for which Poke is a prime platform. It is ray tracer agnostic and can be studied by anyone with access to Python 3.8+. Furthermore, we believe this to be the first open-source implementation of the \emph{transfer-matrix} method of GBD pioneered by Worku and Gross. Harvey et al \cite{Harvey15} report on the \emph{complex ray tracing} method implemented in FRED, which is the first known iteration of the technique. The transfer-matrix method generalizes the problem, enabling the propagation of any analytical solution to the general Collins' integral\cite{Collins:70,ashcraft_inreview}. Consequently, we have the ability to alter the electric field profile of the Gaussian beamlets for higher-fidelity diffraction simulation. 

\subsection{Polarization Ray Tracing}

High-contrast imaging instruments in the next decade will be forced to contend with polarization aberrations\cite{max_gpi_polabs}. The theory of polarization aberrations is described by McGuire and Chipman\cite{McGuire:94_1,McGuire:94_2}, and Sasian\cite{Sasian}. These effects introduce low-order aberrations in even perfectly aligned optical systems, resulting from the polarized interaction of light with material interfaces. Yun\cite{Yun:11_dia,Yun:11_ret} outlined the basics of Polarization Ray Tracing (PRT), an algorithm by which polarization aberrations can be diagnosed. An exhaustive review of the technique is covered in Yun and Chipman, and will not be explained in detail here, but we refer the readers to \emph{Polarized Light and Optical Systems}\cite{chipman2018polarized} for a tutorial. PRT propagates the effects of Fresnel Reflections through a 3D optical system to understand the total vector field behavior.  Most notably, PRT gives us access to the Jones pupil. This is a critical data product for conducting diffraction modeling with consideration for polarization aberrations. PRT has been implemented in commercial design codes (CODE V, Zemax), and POLARIS-M\footnote{\url{https://airyoptics.com/polaris-m-software/}}, which operates in Mathematica. However, there isn’t an apparent package with support in the open source community. To address this need, we built PRT into Poke to bring the technique into Python 3.8+. Users can retain their optical system ray traces in their preferred software, but load the PRT data directly into a Python environment. Using Poke, one can generate a Jones pupil using the following code. In this example, we illustrate the polarization aberrations of a Cassegrain telescope with a fold mirror\cite{Breckenridge15}.

\begin{lstlisting}[language=Python, caption=Generating a Jones pupil is as simple as calling the compute jones pupil method on a Rayfront with ray and material data.]
import poke.plotting as plot
rf.compute_jones_pupil(aloc=np.array([0.,1.,0.]))
plot.jones_pupil(rf)
\end{lstlisting}

\begin{figure}[H]
    \centering
    \includegraphics[width=\textwidth]{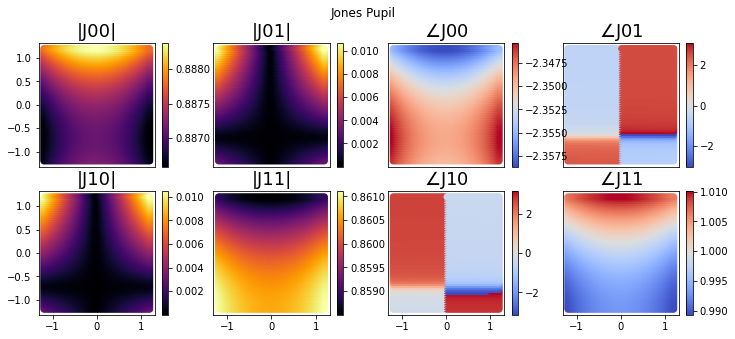}
    \caption{Jones pupil for a Cassegrain with a fold mirror, illustrating the characteristic tilt and astigmatism for the system\cite{Breckenridge15}. The left two columns are the amplitude of the Jones pupil, and the right two are the phase. The on-diagonal elements represent how polarized light transmits through the system, and the off-diagonals represent how one polarization is rotated into the orthogonal component.}
    \label{fig:jones_pupil}
\end{figure}

Poke can decompose the elements of the Jones pupil into Zernike polynomials to re-sample them onto a regularly spaced grid. We can then leverage poke.interfaces to export the data to a physical optics propagator to perform diffraction analysis. 

\begin{lstlisting}[language=Python, caption=Using HCIPy and poke.interfaces to compute the polarized point-spread function of a cassegrain telescope with a fold mirror.]
from hcipy import *
from poke.interfaces import rayfront_to_hcipy_wavefront

# define HCIPy parameters
npix = 256
pupil_grid = make_pupil_grid(npix)
focal_grid = make_focal_grid(8,12)
prop = FraunhoferPropagator(pupil_grid,focal_grid)
telescope_aperture = make_magellan_aperture(True)(pupil_grid)

# convert the Rayfront's jones pupil to an HCIPy wavefront
wavefront = rayfront_to_hcipy_wavefront(rf,npix,pupil_grid)

# apply aperture and propagate
wavefront.electric_field *= telescope_aperture
focused_wavefront = prop(wavefront)
\end{lstlisting}

\begin{figure}[H]
    \centering
    \includegraphics[width=0.5\textwidth]{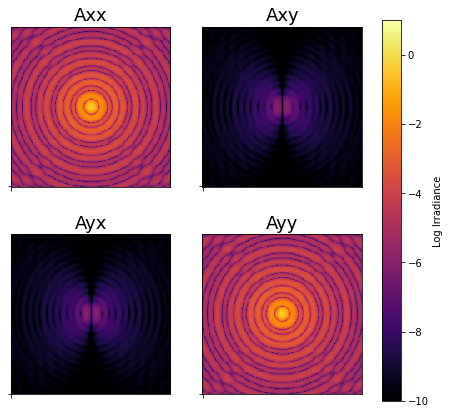}
    \caption{Plot of the amplitude response matrix generated by the previous code Listing 6. The Magellan-clay telescope's aperture was applied in this example. The off-diagonal elements show the contributions of polarization aberrations as "crosstalk" terms that broaden the PSF core.}
    \label{fig:point-spread matrix}
\end{figure}

Poke provides us with the platform we need to democratize PRT, while still being useful for investigators with access to commercial ray tracing licenses. 

\subsection{Homogeneous Thin Film Stacks}
Corollary to PRT is the need to simulate thin film multilayer stacks. Polarization in optical systems is frequently caused by the thin films deposited on the optics. To accurately model these effects in PRT, we built \verb|poke.thinfilms| to be able to encompass the influence of arbitrary stacks of homogeneous, linear, and isotropic thin films. We do so via the characteristic matrix\cite{BYUOpticsBook}, but shown below in Equations \ref{eq:char_p} and \ref{eq:char_s}

\begin{equation}
    A^{(p)} = \frac{1}{2n_{o}cos(\theta_{o})}
    \begin{pmatrix}
        n_{o} & cos(\theta_{o}) \\
        n_{o} & -cos(\theta_{o}) \\
    \end{pmatrix}
    \prod_{j=1}^{N} M_{j}^{(p)}
    \begin{pmatrix}
        cos(\theta_{N+1}) & 0 \\
        n_{N+1} & 0 \\
    \end{pmatrix}
    \label{eq:char_p}
\end{equation}

\begin{equation}
    A^{(s)} = \frac{1}{2n_{o}cos(\theta_{o})}
    \begin{pmatrix}
        n_{o}cos(\theta_{o}) & 1 \\
        n_{o}cos(\theta_{o}) & -1 \\
    \end{pmatrix}
    \prod_{j=1}^{N} M_{j}^{(s)}
    \begin{pmatrix}
        1 & 0 \\
        n_{N+1}cos(\theta_{N+1}) & 0 \\
    \end{pmatrix}
    \label{eq:char_s}
\end{equation}

\begin{figure}[H]
    \centering
    \includegraphics[width=0.8\textwidth]{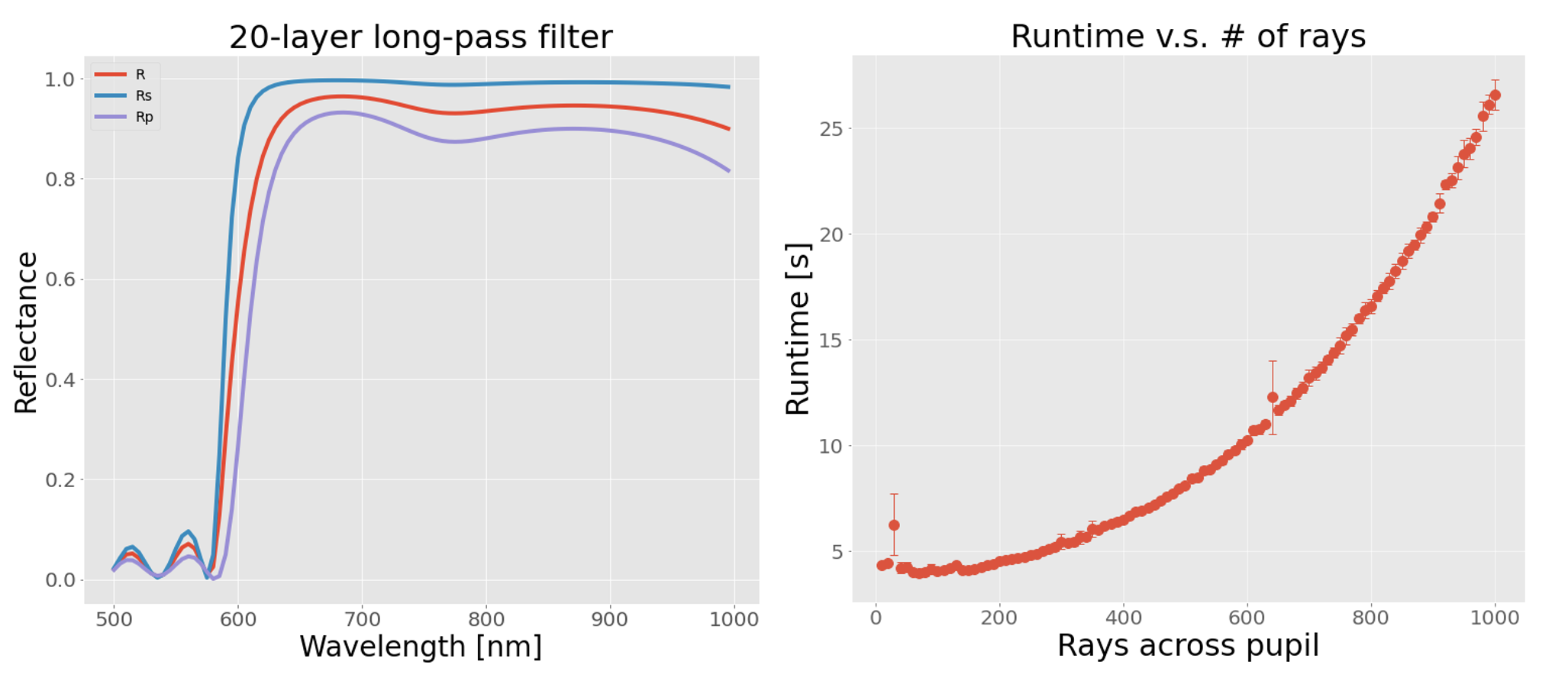}
    \caption{(Left) 20-layer shortpass filter designed using poke.thinfilms and scipy.optimize.minmize. (Right) Runtime to compute a Jones pupil for the 20-layer stack v.s. number of rays across the pupil. There's an $\approx$5 second overhead to initialize the connection to Zemax. The Jones pupil computation on a $64 \times 64$ ray grid takes $\approx$ 10 miliseconds.}
    \label{fig:multilayer_films}
\end{figure}

Equations \ref{eq:char_p} and \ref{eq:char_s} show the characteristic matrix $A$ from which the complex amplitude coefficients are readily derived. While this utility is useful in PRT simulations, it can be called independently to design thin film multilayer stacks. Below we show the design of a long pass filter using \verb|poke.thinfilms| and \verb|scipy.minimize|. This is an excellent example of the utility available to us by bringing these physical optics techniques into Python. In Figure \ref{fig:multilayer_films} we illustrate a short-pass filter that was designed using the thin films module, and the runtime to compute a Jones pupil from this film data.

\begin{figure}[H]
    \centering
    \includegraphics[width=\textwidth]{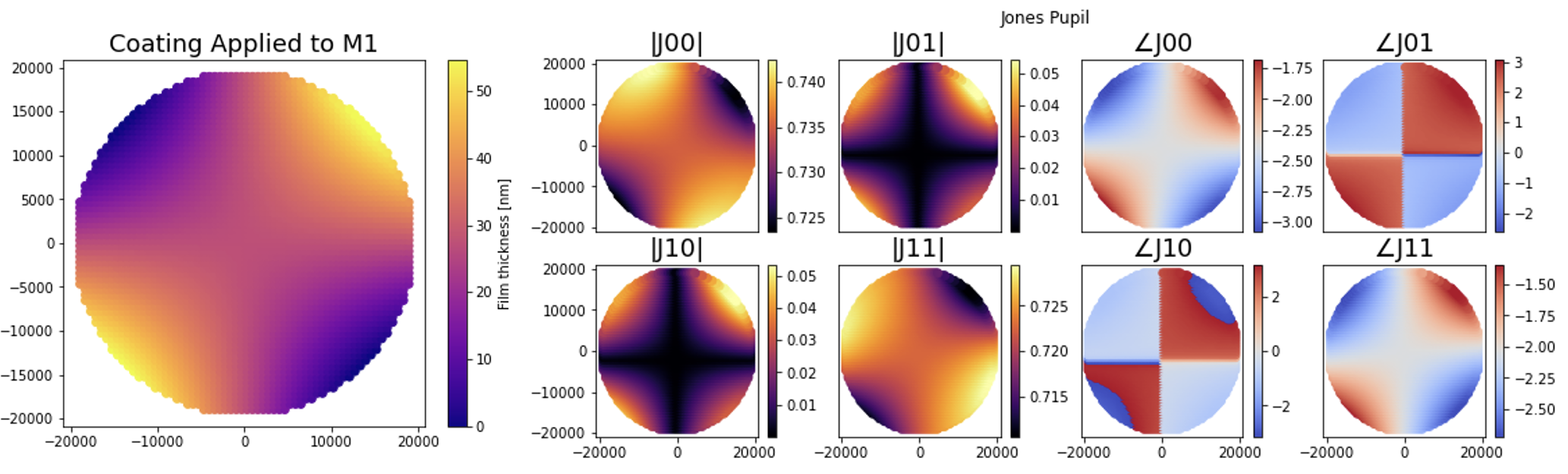}
    \caption{(Left) Astigmatic coating applied to the primary mirror of the ELT. (Right) Jones pupil at $\lambda=600nm$ for the system with an astigmatic coating on M1 and a $\lambda/8$ coating on M2 - M5. The astigmatism appears in both amplitude and phase, largely dominating over the presence of polarization aberrations in the system.}
    \label{fig:elt_astig}
\end{figure}

The thin film interface with optical surfaces is also compatible with spatially-varying thin films. This utility is critical in understanding how changes in film thickness can influence the Jones pupil of an optical system. In Figure \ref{fig:elt_astig} we apply an $45^{o}$ astigmatic coating to the primary mirror of the Extremely Large Telescope (ELT) and evaluate the Jones pupil. From these data, we are able to understand such a film's influence on the amplitude and phase of our polarization state.
 
\section{Integrated Modeling for Future Telescopes}
\label{sec:poke_integrated}
Poke was inspired by the need to create a more integrated modeling pipeline for next-generation astronomical telescopes. Its capabilities to be an interface between commercial ray tracing codes and diffraction modeling software, while adding other ray-based utilities to the modeling pipeline, makes it a powerful design platform. Next-generation astronomical observatories include the 30m class ground-based Giant Segmented-Mirror Telescopes (GSMT's) and the Astro2020-recommended Habitable Worlds Observatory (HWO). 

\subsection{Polarization Effect Modeling}
Poke has already been used to model the nominal polarization aberrations present in the GSMT's. In Anche et al \cite{Anche2023} we discovered that the polarization aberrations due to the extremely large angles of incidence on the observatory mirrors lead to polarization aberration residuals that were greater than the desired contrast at the focal plane of ideal coronagraphic instrumentation. To achieve the baseline contrast proposed for these instruments, we must develop a method to mitigate the effect of polarization aberrations. In Ashcraft et al. in prep, we are developing an optimization-based method to design thin-film stacks to mitigate polarization aberrations. 

\begin{figure}[H]
    \centering
    \includegraphics[width=0.7\textwidth]{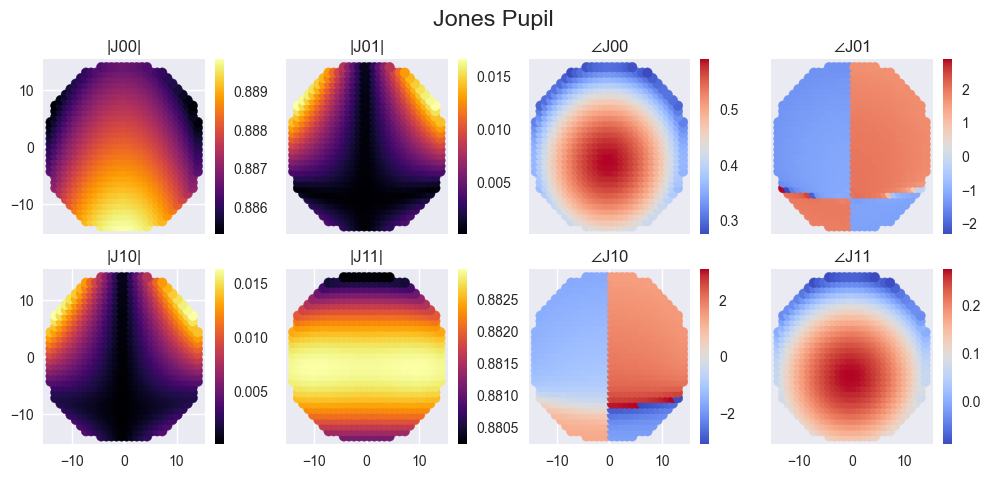}
    \caption{Figure showing the Jones pupil of the ELT. The left two columns represent the amplitude of the complex pupil, and the right two represent the phase. There is an apparent difference in focus-like phase aberrations between the on-diagonals of the phase quadrant (J00, J11).}
    \label{fig:ground_600nm}
\end{figure}

\begin{figure}[H]
    \centering
    \includegraphics[width=0.7\textwidth]{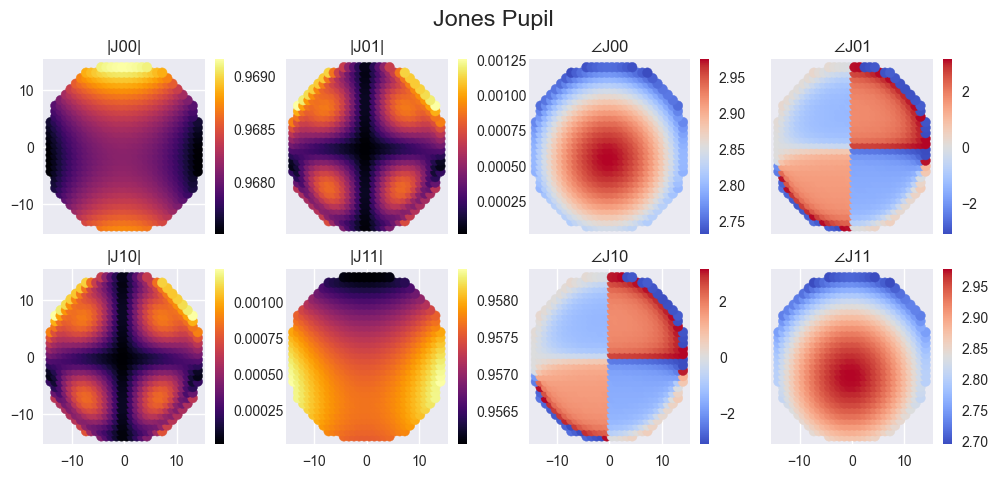}
    \caption{The Jones pupil of the ELT at $\lambda=600nm$ after optimization of the thin films of each optic in the telescope. The throughput (on-diagonal terms, left two columns) is 7-8$\%$ higher, and the cross-talk (off-diagonal terms, left two columns) decreased by almost an order of magnitude when compared to the result in Figure \ref{fig:ground_600nm}. The phase aberrations of each polarization state (on-diagonal terms, right two columns) are also nearly identical, which indicates that a deformable mirror can better null the difference between the polarization states.}
    \label{fig:jones_after}
\end{figure}

In Figures \ref{fig:ground_600nm}-\ref{fig:jones_after} we illustrate the results of using PRT with optimization routines in Python to design thin film stacks that can reduce the presence of polarization aberrations. This technique has potential implications for maximizing the achievable contrast near the inner working angle of coronagraphs for next-generation astronomical observatories. 

\subsection{High-performance Computing Options}
Poke defaults to \verb|numpy| for the vast majority of its numerical calculations. However, for high-fidelity simulations it is advantageous to employ Python packages with \verb|numpy|-like API's that grant the user more utility. Poke has adopted the interchangeable backend system of prysm (see section 4A of Dube et al\cite{Dube2022}). Poke's backend can be swapped from \verb|numpy| to \verb|cupy| for accelerated GPU-based computation, and \verb|jax| to enable automatic differentiation.

\verb|cupy| is an incredible Python package that allows users to scale their computations to GPUs without making significant changes to their code. It's numpy-like API makes it trivial to implement in any codebase. GPUs are especially suited to large matrix multiplications, which are the fundamental operation of the physical optics modules discussed in Section 3.

\verb|jax| is a Python package developed by Google that supports automatic differentiation (or autodiff). Autodiff enables the computation of analytic gradients without requiring the user to derive the analytic expression for gradient back-propagation. This is particularly useful for complex expressions, where the mathematics would grow unwieldy. \verb|jax| makes Poke fully differentiable, enabling accurate computation of gradients for accelerated optimization. 

\section{Conclusions and Future Work}
\label{sec:discussion}
Poke is a new open-source ray-based physical optics engine written to further integrate the optical modeling pipeline. The availability of ray data in a Python environment democratizes the analysis done by ray tracing engines. The physics modules developed for Poke enhance the user's simulation capabilities. This proceedings presents a brief overview of the utilities of Poke and their implications for modeling and design efforts in support of future astronomical observatories.

We are actively searching for collaborators to help build Poke. If you are interested in contributing, please contact the author or open an issue on our Github page\footnote{https://github.com/Jashcraf/poke}\cite{jaren_ashcraft_2023_7951685} to start a discussion. Areas we are currently seeking to improve include:

\begin{enumerate}
    \item Interface with an open-source ray tracer
    \item Simulating the complex amplitude coefficients of liquid crystal thin films
    \item Simulating the complex amplitude coefficients of metasurfaces
    \item Exploring alternative beamlet profiles for more accurate beamlet decomposition techniques
    \item Contribution to examples in \url{poke.readthedocs.io}
    \item An active interface with commercial ray tracers for real-time interaction with optical system ray traces.
\end{enumerate}



\acknowledgments 
Kian Milani and Kevin Derby contributed to the GPU compatibility. Brandon Dube wrote the function to read/write Rayfronts. Emory Jenkins wrote the Zernike polynomial generator. Ramya Anche and Trent Brendel contributed tests to the PRT and GBD modules, respectively. This work was funded by an NASA Space Technology Graduate Research Opportunity.

\bibliography{report} 

\begin{thebibliography}{10}

\bibitem{2016ascl.soft02018P}
{Perrin}, M., {Long}, J., {Douglas}, E., {Sivaramakrishnan}, A., {Slocum}, C.,
  and {others}, ``{POPPY: Physical Optics Propagation in PYthon}.''
  Astrophysics Source Code Library, record ascl:1602.018 (Feb. 2016).

\bibitem{HCIPYdocs}
Por, E.~H., Haffert, S.~Y., Radhakrishnan, V.~M., Doelman, D.~S., van Kooten,
  M., and Bos, S.~P., ``{High Contrast Imaging for Python (HCIPy): an
  open-source adaptive optics and coronagraph simulator},'' in [{\em Adaptive
  Optics Systems VI}{\nolinebreak\hspace{0.1em}]},  Close, L.~M., Schreiber,
  L., and Schmidt, D., eds.,  {\bf 10703},  1112 -- 1125, International Society
  for Optics and Photonics, SPIE (2018).

\bibitem{Dube2019}
Dube, B., ``prysm: A python optics module,'' {\em Journal of Open Source
  Software}~{\bf 4}(37),  1352 (2019).

\bibitem{jaren_ashcraft_2023_7951685}
Ashcraft, J. and hilltailor, ``Jashcraf/poke: v1.0.1,'' (May 2023).

\bibitem{Anche2023}
Anche, R.~M., Ashcraft, J.~N., Haffert, S.~Y., Millar-Blanchaer, M.~A.,
  Douglas, E.~S., Snik, F., Williams, G., van Holstein, R.~G., Doelman, D.,
  Gorkom, K.~V., and Skidmore, W., ``Polarization aberrations in
  next-generation giant segmented mirror telescopes ({GSMTs}),'' {\em Astronomy
 and ~{\bf 672},  A121 (apr 2023).

\bibitem{ashcraft_inreview}
Ashcraft, J.~N., Douglas, E.~S., Kim, D., and Riggs, A., ``Hybrid propagation
  physics for the design and modeling of astronomical observatories: A
  coronagraphic example,'' {\em SPIE Journal of Astronomical Telescopes,
  Instruments, and Systems}  (\emph{In Review}).

\bibitem{Harvey15}
Harvey, J.~E., Irvin, R.~G., and Pfisterer, R.~N., ``{Modeling physical optics
  phenomena by complex ray tracing},'' {\em Optical Engineering}~{\bf 54}(3),
  1 -- 12 (2015).

\bibitem{Worku19}
Worku, N.~G. and Gross, H., ``Propagation of truncated gaussian beams and their
  application in modeling sharp-edge diffraction,'' {\em J. Opt. Soc. Am.
  A}~{\bf 36},  859--868 (May 2019).

\bibitem{Greynolds86}
Greynolds, A.~W., ``{Vector Formulation Of The Ray-Equivalent Method For
  General Gaussian Beam Propagation},'' in [{\em Current Developments in
  Optical Engineering and Diffraction Phenomena}{\nolinebreak\hspace{0.1em}]},
  Fischer, R.~E., Harvey, J.~E., and Smith, W.~J., eds.,  {\bf 0679},  129 --
  133, International Society for Optics and Photonics, SPIE (1986).

\bibitem{Collins:70}
Collins, S.~A., ``Lens-system diffraction integral written in terms of matrix
  optics$\ast$,'' {\em J. Opt. Soc. Am.}~{\bf 60},  1168--1177 (Sep 1970).

\bibitem{max_gpi_polabs}
Millar-Blanchaer, M.~A., Anche, R.~M., Nguyen, M.~M., and Maire, J., ``{The
  polarization aberrations of the Gemini Telescope as seen by the Gemini Planet
  Imager},'' in [{\em Ground-based and Airborne Instrumentation for Astronomy
  IX}{\nolinebreak\hspace{0.1em}]},  Evans, C.~J., Bryant, J.~J., and Motohara,
  K., eds.,  {\bf 12184},  121843X, International Society for Optics and
  Photonics, SPIE (2022).

\bibitem{McGuire:94_1}
McGuire, J.~P. and Chipman, R.~A., ``Polarization aberrations. 1. rotationally
  symmetric optical systems,'' {\em Appl. Opt.}~{\bf 33},  5080--5100 (Aug
  1994).

\bibitem{McGuire:94_2}
McGuire, J.~P. and Chipman, R.~A., ``Polarization aberrations. 2. tilted and
  decentered optical systems,'' {\em Appl. Opt.}~{\bf 33},  5101--5107 (Aug
  1994).

\bibitem{Sasian}
Sasi{\'a}n, J., ``{Polarization fields and wavefronts of two sheets for
  understanding polarization aberrations in optical imaging systems},'' {\em
  Optical Engineering}~{\bf 53}(3),  035102 (2014).

\bibitem{Yun:11_dia}
Yun, G., Crabtree, K., and Chipman, R.~A., ``Three-dimensional polarization
  ray-tracing calculus i: definition and diattenuation,'' {\em Appl. Opt.}~{\bf
  50},  2855--2865 (Jun 2011).

\bibitem{Yun:11_ret}
Yun, G., McClain, S.~C., and Chipman, R.~A., ``Three-dimensional polarization
  ray-tracing calculus ii: retardance,'' {\em Appl. Opt.}~{\bf 50},  2866--2874
  (Jun 2011).

\bibitem{chipman2018polarized}
Chipman, R.~A., Lam, W.-S.~T., and Young, G.,  [{\em Polarized light and
  optical systems}{\nolinebreak\hspace{0.1em}]}, CRC press (2018).

\bibitem{Breckenridge15}
Breckinridge, J.~B., Lam, W. S.~T., and Chipman, R.~A., ``Polarization
  aberrations in astronomical telescopes: The point spread function,'' {\em
  Publications of the Astronomical Society of the Pacific}~{\bf 127}(951),
  445--468 (2015).

\bibitem{BYUOpticsBook}
Peatross, J. and Ware, M.,  [{\em Physics of Light and
  Optics}{\nolinebreak\hspace{0.1em}]}, optics.byu.edu (2015).

\bibitem{Dube2022}
{Dube}, B.~D., {Riggs}, A.~J., {Kern}, B.~D., {Cady}, E.~J., {Krist}, J.~E.,
  {Zhou}, H., {Nemati}, B., {Seo}, B.-J., {Steeves}, J., {Arndt}, D.,
  {Mandi{\'c}}, M., {Shields}, J., {Boussalis}, D., {Valverde}, A., {Rahman},
  Z., and {Fathpour}, N., ``{Exascale integrated modeling of low-order
  wavefront sensing and control for the Roman Coronagraph instrument},'' {\em
  Journal of the Optical Society of America A}~{\bf 39},  C133 (Dec. 2022).}

\end{thebibliography}
\bibliographystyle{spiebib} 

\end{document}